\documentclass[preprint,5p,times,twocolumn]{elsarticle}

\usepackage[latin1]{inputenc}                    
\usepackage{graphicx}                            
\usepackage{latexsym}                            
\usepackage{amsfonts}                            
\usepackage{amssymb}                             
\usepackage{amsmath}                             
\usepackage[mathscr]{eucal}                      
\usepackage{dcolumn}                             
\usepackage{theorem}                             
\usepackage{footnote}
\usepackage{enumitem}
\usepackage{bm}
\usepackage{color}
\usepackage{natbib}

\def\be{\begin{equation}}
\def\ee{\end{equation}}
\def\bea{\begin{eqnarray}}
\def\eea{\end{eqnarray}}

\journal{Physics Letters B}

\begin{document}

\begin{frontmatter}

\title{Electromagnetic contribution to charge symmetry violation in parton distributions}

\author{X.G.~Wang}
 
\author{A.~W.~Thomas}

\author{R.~D.~Young}

\address{\mbox{ARC Centre of Excellence for Particle Physics
	at the Terascale and CSSM}, Department of Physics,\\
	University of Adelaide, Adelaide SA 5005, Australia}

\date{\today}

\begin{abstract}
We report a calculation of the combined effect of photon radiation and 
quark mass differences on charge symmetry violation (CSV) in the parton 
distribution functions of the nucleon. 
{}Following a recent suggestion of Martin 
and Ryskin, the initial photon distribution is calculated in terms
of coherent radiation from the proton as a whole, while the effect of the 
quark mass difference is based on a recent lattice QCD simulation. The distributions 
are then evolved to a scale at which they can be compared with 
experiment by including both QCD and QED radiation.  
Overall, at a scale of 5 GeV$^2$,  
the total CSV effect on the phenomenologically important 
difference between the $d$ and  
$u$-quark distributions is some 20\% larger than the value based on  
quark mass differences alone.
In total these sources of CSV account for approximately 40\% of the 
NuTeV anomaly.
\end{abstract}


\end{frontmatter}

\section{Introduction}
Charge symmetry (CS) refers to the invariance of the QCD Hamiltonian under 
the operator $e^{i \pi I_2}$, a rotation by $180\deg$ about the 2-axis 
in iso-space. Under this operation $u$-quarks rotate into $d$-quarks and 
vice-versa, while protons and neutrons are also interchanged. As a result, 
if QCD were to respect this symmetry, the up-quark distribution in the proton, 
$u^p$, and the down quark distribution in the neutron, $d^n$, would be 
identical. Similarly one would have $d^p \, = \, u^n$. Precisely these 
relations have been almost universally {\it assumed} for the past 40 years, 
as without such an assumption it would have been impossible to separate 
the flavor dependence of the parton distribution functions (PDFs). 

Studies of strongly interacting systems have established that CS is 
typically respected at the level of a fraction of a 
percent~\cite{Miller:1990iz}, much 
better than isospin symmetry, which requires the invariance of the 
Hamiltonian under all rotations in iso-space. Nevertheless, as one 
uses tests of symmetries to probe for physics beyond the Standard Model,  
or aims for higher precision at the LHC, it is vital to know just 
how well the PDFs satisfy CS. Furthermore, subtle tests of such 
a symmetry can also yield information on how QCD itself works. Thus the 
study of charge symmetry violation (CSV) in PDFs may also lead to a deeper 
understanding of the structure of the nucleon itself. For reviews 
of CSV in PDFs we refer to Londergan 
{\it et al.}~\cite{Londergan:2009kj,Londergan:1998ai}.

There are two dominant sources of CSV in the nucleon, the electromagnetic 
interaction and the mass differences of the $u$ and $d$ quarks, 
$\delta m \, = \, m_d \, - \, m_u$. The first investigations of CSV 
in the PDFs were based on the effect of $\delta m$ within the MIT 
bag model~\cite{Sather:1991je,Rodionov:1994cg,Londergan:1994gr}. These 
calculations showed CSV violating effects as large as 5\% at large-$x$, 
while they were at the percent level in the momentum fractions:
\begin{equation}
\label{eq:csvPDFs}
\delta U \, = \, \int_0^1 dx \, x \delta u(x) \, \, ; \, \, 
\delta D \, = \, \int_0^1 dx \, x \delta d(x) \, \, ,
\end{equation}
where the CSV PDFs are $\delta u \, = \, u^p - d^n$ and $\delta d \, = \, 
d^p - u^n$. The major effect, which could be understood in terms of the 
dominant role played by di-quark correlations~\cite{Close:1988br}, 
arose from the mass difference between the $dd$ and $uu$ spectator 
pairs to the struck $u$-quark in a neutron and $d$-quark in a proton.
It was found that $\delta u$ and $\delta d$ had a similar magnitude 
and opposite sign.

In the context of the NuTeV experiment~\cite{Zeller:2001hh}, 
where these published effects 
were sufficiently large to reduce the anomaly to 2$\sigma$ 
or less~\cite{Londergan:2003ij,Londergan:2003pq} considerable work was 
carried out to establish the extent to which these results were 
model independent. Recently, lattice QCD studies of these 
moments~\cite{Horsley:2010th,Cloet:2012db}
(although necessarily the charge conjugation positive combination, rather 
than the valence combination calculated in the bag model) confirmed 
the sign and magnitude of the pioneering calculations. 

The importance of QED radiation on DIS processes was recognised more 
than 40 years ago in the context of photon radiation from quarks in 
charged current neutrino interactions~\cite{Kiskis:1973ia}. In the 
context of DGLAP evolution of PDFs, Spiesberger~\cite{Spiesberger:1994dm}
summarised the potentially large effects associated with mass singularities 
involving $\ln (Q_0^2/m_q^2)$, where $m_q$ is a light quark current mass.
To avoid such problems he proposed to redefine the PDFs at the starting 
scale, $Q_0^2$, to include the effects of these singularities.
The residual effects of photon radiation are then relatively small 
and at any given scale, $Q^2$, could be shown to be equivalent to a shift 
of scale of the PDFs by a charge-dependent factor (analogous to 
``dynamical rescaling'').

In the modern era, Martin {\it et al.}[MRST]~\cite{Martin:2005sk} and 
Gl\"uck {\it et al.}~\cite{Gluck:2005xh} sought to improve on the 
work of Spiesberger by including a photon distribution at the starting 
scale. In both cases this meant that CS was violated at that scale and 
both the initial photon distribution and the CSV PDFs were estimated in 
terms of the large logarithms associated with the mass singularities.
Again in the context of the NuTeV experiment, it is important that the 
sign of the CSV associated with QED radiation was the {\it same} as that 
arising from $\delta m$~\cite{Gluck:2005xh}, 
even though they enter with opposite signs 
in the neutron-proton mass difference. 
It is therefore vital~\cite{Bentz:2009yy} to have 
a consistent treatment of both effects and this is the aim of the 
present work.

The appearance of current masses in the QED logarithm is at odds with
the modern understanding of non-perturbative QCD. At low scales
the phenomenon of spontaneous chiral symmetry
breaking~\cite{Thomas:2001kw} means that what naturally appears is a
constituent quark mass, rather than the current quark mass. For
example, a naive evaluation of the electromagnetic self-energy of a
quark including the non-perturbative quark propagator, naturally
yields a result proportional to $e_q^2 \alpha M(0)$, where $M(0) \sim 0.4$
GeV~\cite{Bashir:2012fs} and $e_q$ is the charge of the quark in units
of the positron charge. Since the quark-photon splitting function is
derived by cutting this self-energy diagram, one is rather led to a
correction to the PDFs at the starting scale of order $\ln
(Q_0^2/M(0)^2)$, which is necessarily much smaller than proposed in
Refs.~\cite{Martin:2005sk,Gluck:2005xh}.

Very recently Martin and Ryskin~\cite{Martin:2014nqa} re-examined the 
issue of the initial photon distribution. They observed that at the 
low scale $Q_0^2$ the major part of the input photon distribution 
``comes from the {\it coherent} emission of the photon from the 
`elastic' proton''. In the present work we use this insight to 
make a new and more consistent calculation of the electromagnetic 
contributions to CSV. 
Since the scale at which typical, valence dominated quark models like 
NJL are matched to QCD is somewhat lower that that used in 
Refs.~\cite{Martin:2005sk,Gluck:2005xh}, we modify the initial quark 
distribution and generate the initial photon distribution for the 
coherent radiation from the elastic proton. The evolution from that 
scale to a typical scale at which one might compare with experiment, 
say 5 GeV$^2$, then includes incoherent radiation of both gluons and 
photons from the quarks. 
In comparison
with simply adding the effect of the quark mass
difference to the original estimates of QED radiation by MRST and 
Gl\"uck {\it et al.}, the extent of
CSV on the $u$-quarks increases a little, while that for the $d$-quarks
decreases in magnitude. Overall, at a scale of 5 GeV$^2$,
the total CSV effect on the difference between the $d$ and
$u$-quark distributions, which is the combination relevant to the NuTeV 
experiment, is some 20\% larger than the value based on
quark mass differences alone.

\section{Quark distribution functions}
\label{sec:GRV}

The dynamical parton distributions, generated radiatively from valence-like inputs at low scales, are determined from global fit by Gl\"uck, Reya and Vogt (GRV)~\cite{Gluck:1998}, 
taking into account small-$x$ data on deep inelastic and other hard scattering processes.
The leading order (LO) input distributions of 
proton at $Q_0^2=\mu_{\rm LO}^2=0.26$ GeV$^2$ are 
then given by
\begin{eqnarray}
xu_v(x,\mu_{\rm LO}^2) & = & 1.239 x^{0.48} (1-x)^{2.72}\nonumber\\
&&    \times (1-1.8\sqrt{x} + 9.5x)\nonumber\\
xd_v(x,\mu_{\rm LO}^2) & = & 0.614 (1-x)^{0.9} xu_v(x,\mu_{\rm LO}^2)\nonumber\\
x\Delta(x,\mu_{\rm LO}^2) & = & 0.23 x^{0.48} (1-x)^{11.3}\nonumber\\
&&   \times (1-12.0\sqrt{x} + 50.9x)\nonumber\\
x(\bar{u}+\bar{d})(x,\mu_{\rm LO}^2) & = & 1.52 x^{0.15} (1-x)^{9.1}\nonumber\\
&&   \times (1-3.6\sqrt{x} + 7.8x)\nonumber\\
xg(x,\mu_{\rm LO}^2) & = & 17.47 x^{1.6} (1-x)^{3.8}\nonumber\\
xs(x,\mu_{\rm LO}^2) & = & x\bar{s}(x,\mu_{\rm LO}^2) = 0\ ,
\end{eqnarray}
where $\Delta\equiv\bar{d}-\bar{u}$.  The corresponding
next-to-leading order (NLO) input at $Q_0^2=\mu_{\rm NLO}^2=0.40$ GeV$^2$ 
is
\begin{eqnarray}
xu_v(x,\mu_{\rm NLO}^2) & = & 0.632 x^{0.43} (1-x)^{3.09} (1+18.2x)\nonumber\\
xd_v(x,\mu_{\rm NLO}^2) & = & 0.624 (1-x)^{1.0} xu_v(x,\mu_{\rm NLO}^2)\nonumber\\
x\Delta(x,\mu_{\rm NLO}^2) & = & 0.20 x^{0.43} (1-x)^{12.4}\nonumber\\
&&    \times (1-13.3\sqrt{x} + 60.0x)\nonumber\\
x(\bar{u}+\bar{d})(x,\mu_{\rm NLO}^2) & = & 1.24 x^{0.20} (1-x)^{8.5}\nonumber\\
&&    \times (1-2.3\sqrt{x} + 5.7x)\nonumber\\
xg(x,\mu_{\rm NLO}^2) & = & 20.80 x^{1.6} (1-x)^{4.1}\nonumber\\
xs(x,\mu_{\rm NLO}^2) & = & x\bar{s}(x,\mu_{\rm NLO}^2) = 0.
\end{eqnarray}
%


\section{Photon distribution functions}
\label{sec:initial-photon}
The additional contribution to the valence quark charge asymmetries 
arises from radiative QED effects. The so-called DGLAP evolution equations 
are modified by introducing the photon PDF, 
$\gamma^N(x,Q^2)$. Following Martin and Ryskin,
as shown in Ref.~\cite{Martin:2014nqa}, the major part of the input 
photon PDF, $\gamma^p(x,Q^2_0)$, 
comes from the {\it coherent} emission of the photon from the elastic proton. 
Below the model scale, $Q_0^2$,  we assume that the 
contribution from {\it incoherent}  
emission of photons from quarks 
within the nucleon is negligible.
Above the model scale we utilize the APFEL
program~\cite{APFEL:2014} to perform combined LO/NLO QCD
and LO QED evolution in the variable-flavor-number
scheme (VFNS). That is, in that region both QCD and QED radiation 
is treated as incoherent radiation from the quarks.

The coherent emission from the proton is given by~\cite{Martin:2014nqa}
\begin{equation}\label{eq:gamma-p-coherent}
\gamma^p_{coh}(x,Q^2_0) = \frac{\alpha}{2\pi}\frac{1+(1-x)^2}{x}
\int_0^{|t|<Q_0^2} d q_t^2 \frac{q_t^2}{(q_t^2 + x^2 m_p^2)^2} F_1^2(t)\ ,
\end{equation}
where $q_t$ is the transverse momentum of the emitted photon and
\begin{equation}
t = - \frac{q_t^2 + x^2 m_p^2}{1-x}\ .
\end{equation}
$F_1$ is the Dirac form factor of proton. Letting $Q^2 = -t$, $F_1(Q^2)$ is given by
\begin{equation}
F_1(Q^2) = \frac{4M_p^2 G^p_E(Q^2) + Q^2 G^p_M(Q^2)}{Q^2 + 4M_p^2}\ ,
\end{equation}
with a dipole parametrization for the electric and magnetic form factors,
\begin{equation}
G^p_E(Q^2) = \frac{G^p_M(Q^2)}{\mu_p} = \frac{1}{(1+\frac{Q^2}{\Lambda^2})^2}\ ,
\end{equation}
where $\mu_p = 2.793$ and $\Lambda^2 = 0.71\ \mathrm{GeV}^2$\ .

For the neutron, we neglect the small $F_1$ form factor of the neutron
and set
\begin{equation}
\label{eq:gamma-n-coherent}
\gamma^n_{coh} (x,Q_0^2) = 0\ .
\end{equation}
The proton momentum fraction carried by the photon is
\begin{equation}
p_{\gamma}(Q_0^2) = \int _0^1 x \gamma^p_{coh}(x,Q_0^2) d x\ .
\end{equation}
Correspondingly, the initial distribution functions of the 
valence quarks in proton should be modified as
\begin{eqnarray}\label{eq:quark-coherent}
u_v^p(x,Q_0^2) &=& \left[ u_v^p(x,Q_0^2) \right]_{\mathrm{GRV}} 
- \beta^u f(x,Q_0^2)\ ,\nonumber\\
d_v^p(x,Q_0^2) &=& \left[ d_v^p(x,Q_0^2) \right]_{\mathrm{GRV}} 
- \beta^d f(x,Q_0^2)\ ,
\end{eqnarray}
where $f(x,Q^2)$ at $Q^2 = 1\ \mathrm{GeV}^2$ is taken from Ref.~\cite{Martin:2003sk},
\begin{equation}
f(x,Q^2=1\,{\rm GeV}^2) = x^{-0.5} (x-0.0909)(1-x)^4\ .
\end{equation}
The quantity $f(x)$ is chosen since its $x$ dependence has 
roughly the same form as the MRST initial valence quark parton 
distribution functions in both the limit $x \rightarrow 0$ 
and $x \rightarrow 1$ at $Q^2 = 1\ \mathrm{GeV}^2$. 
The first moment of $f(x)$ is fixed to be zero, in agreement with the valence quark normalization.
$f(x,Q_0^2)$ at $Q_0^2 = 0.26\ \mathrm{GeV}^2$ and 
$Q_0^2 = 0.40\ \mathrm{GeV}^2$ are obtained by LO and NLO QCD
evolution from $Q^2 = 1\,\mathrm{GeV}^2$, respectively.

The coefficients $\beta^u$ and $\beta^d$ are determined by 
assuming that the momentum loss of the valence $u$ and $d$ quarks 
in the proton are $\frac{2}{3}p_{\gamma}$ and
$\frac{1}{3}p_{\gamma}$, respectively, 
\begin{equation}
\beta^u \int_0^1 dx \, x f(x,Q_0^2) 
= 2 \beta^d \int_0^1 dx \, x f(x,Q_0^2) = \frac{2}{3} p_{\gamma}(Q_0^2)\ .
\end{equation}
The momentum fraction carried by the photon and the corresponding 
$\beta$ parameters are shown in Table~\ref{tab:parameter-beta}.

\begin{table}[ht]
\begin{center}
\caption{\label{tab:parameter-beta} The momentum fraction carried by the photon and the corresponding $\beta$ parameters.}
\renewcommand{\arraystretch}{1.3}
\begin{tabular}{c|c|cc}\hline
$Q_0^2\ (\mathrm{GeV}^2)$   &   $p_{\gamma}(Q_0^2)$  &    $\beta^u$   &     $\beta^d$ \\ \hline
  0.26        &               0.00105              &      0.0584      &        0.0292  \\ 
  0.40         &              0.00113              &      0.0614      &        0.0307  \\ \hline    
\end{tabular}
\end{center}
\end{table}

\section{Results}
\begin{figure}[h]
\begin{center}
\includegraphics[width=8cm]{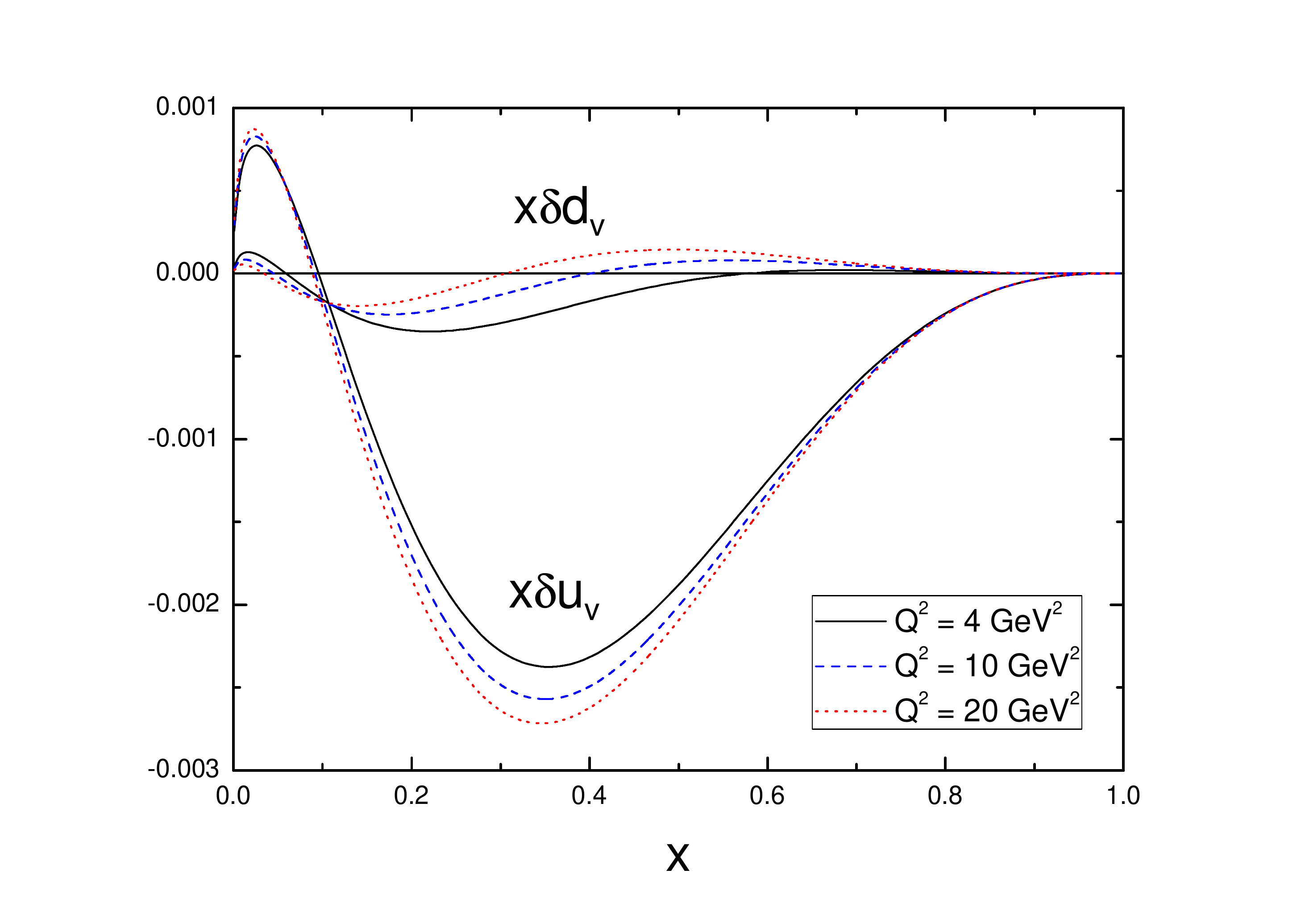}\\
\includegraphics[width=8cm]{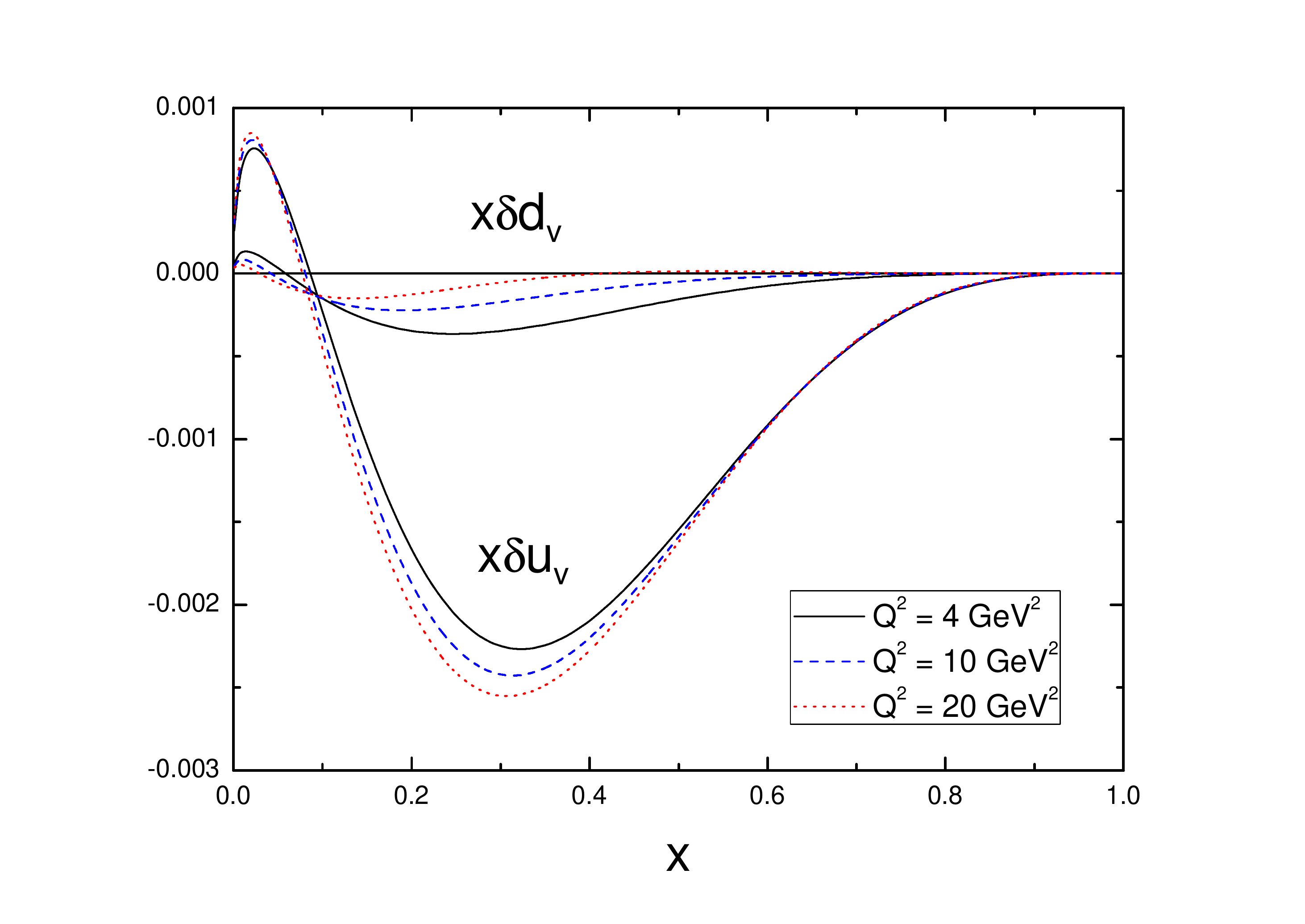}
\caption{(colour online). The pure QED contributions to isospin-violating majority $x\delta u_v$ 
and minority $x\delta d_v$ valence parton distributions 
at $Q^2=4\ \mathrm{GeV}^2$, $10\ \mathrm{GeV}^2$ and $20\ \mathrm{GeV}^2$. 
(Upper) LO QCD evolution from LO initial distributions at $Q_0^2=0.26\ \mathrm{GeV}^2$; 
(Lower) NLO QCD evolution from NLO initial distributions at $Q_0^2=0.40\ \mathrm{GeV}^2$.
In both cases, the QED evolution is of LO in $\alpha$.}
\label{fig:csv-coherent}
\end{center}
\end{figure}

In each case, we evolve to the final scales $Q^2 = 4\ \mathrm{GeV}^2$, $10\ \mathrm{GeV}^2$ and $20\ \mathrm{GeV}^2$.
The pure QED contributions to the isospin-violating majority and minority valence distributions, 
$x\delta u_v$ and $x\delta d_v$, respectively,  
are shown in Fig.~\ref{fig:csv-coherent}. 
The corresponding contributions to the second moments of $\delta u_v$ 
and $\delta d_v$ are given in Table~\ref{tab:delta-Q-v-coherent}.

\begin{table*}[!htbp]
\begin{center}
\caption{\label{tab:delta-Q-v-coherent} The pure QED contributions to the second moments of $\delta u_v$ 
and $\delta d_v$ at $Q^2 = 4\ \mathrm{GeV}^2$, $10\ \mathrm{GeV}^2$ and $20\ \mathrm{GeV}^2$. 
The photon and valence quark distribution functions at the initial scale 
are given by Eqs.~(\ref{eq:gamma-p-coherent}), (\ref{eq:gamma-n-coherent}) 
and (\ref{eq:quark-coherent}), respectively.}
\renewcommand{\arraystretch}{1.3}
\begin{tabular}{c|cc|cc|rr}\hline
                   $Q_0^2$             & \multicolumn{2}{c|}{$Q^2=4\ \mathrm{GeV}^2$}  & \multicolumn{2}{c|}{$Q^2=10\ \mathrm{GeV}^2$} & \multicolumn{2}{c}{$Q^2=20\ \mathrm{GeV}^2$}  \\ \cline{2-7}
         $(\mathrm{GeV}^2)$    &            $\delta U_v$        &       $\delta D_v$         &               $\delta U_v$       &      $\delta D_v$         &               $\delta U_v$       &      $\delta D_v$        \\ \hline
                      0.26                 &            $-0.00099$          &        $-0.00009$          &               $-0.00107$         &       $-0.00003$          &               $-0.00113$         &       $0.00001$            \\
                      0.40                 &            $-0.00089$          &        $-0.00013$          &               $-0.00095$         &       $-0.00007$          &               $-0.00099$         &      $-0.00003$            \\ \hline
\end{tabular}
\end{center}
\end{table*}

At the initial scale, both $\delta U_v$ and $\delta D_v$ are negative
in the valence region. At higher scales, $\delta U_v$ will decrease
and therefore always remain negative.  The sign of $\delta D_v$
depends on the final scale $Q^2$.  There exists a critical scale,
$Q_c^2$, above which $\delta D_v$ will be positive.
At $Q^2 = 10\ \mathrm{GeV}^2$, which is appropriate for the NuTeV 
experiment, the QED contributions to the second moments for both the $u$ 
and $d$ quarks are significantly smaller than the predictions of Refs.~ \cite{Martin:2005sk} and \cite{Gluck:2005xh}.

\begin{figure}[h]
\begin{center}
\includegraphics[width=8cm]{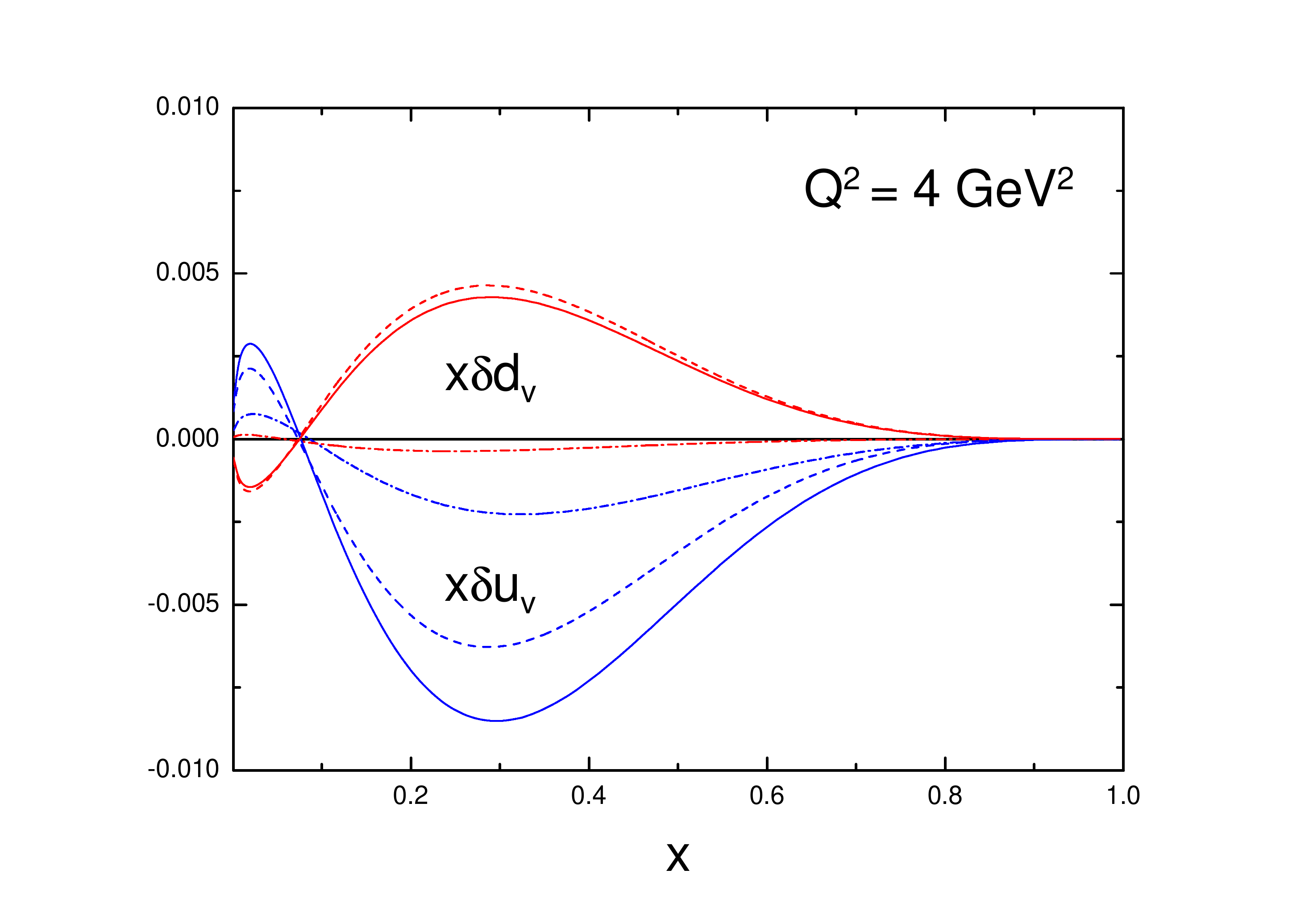}\\
\includegraphics[width=8cm]{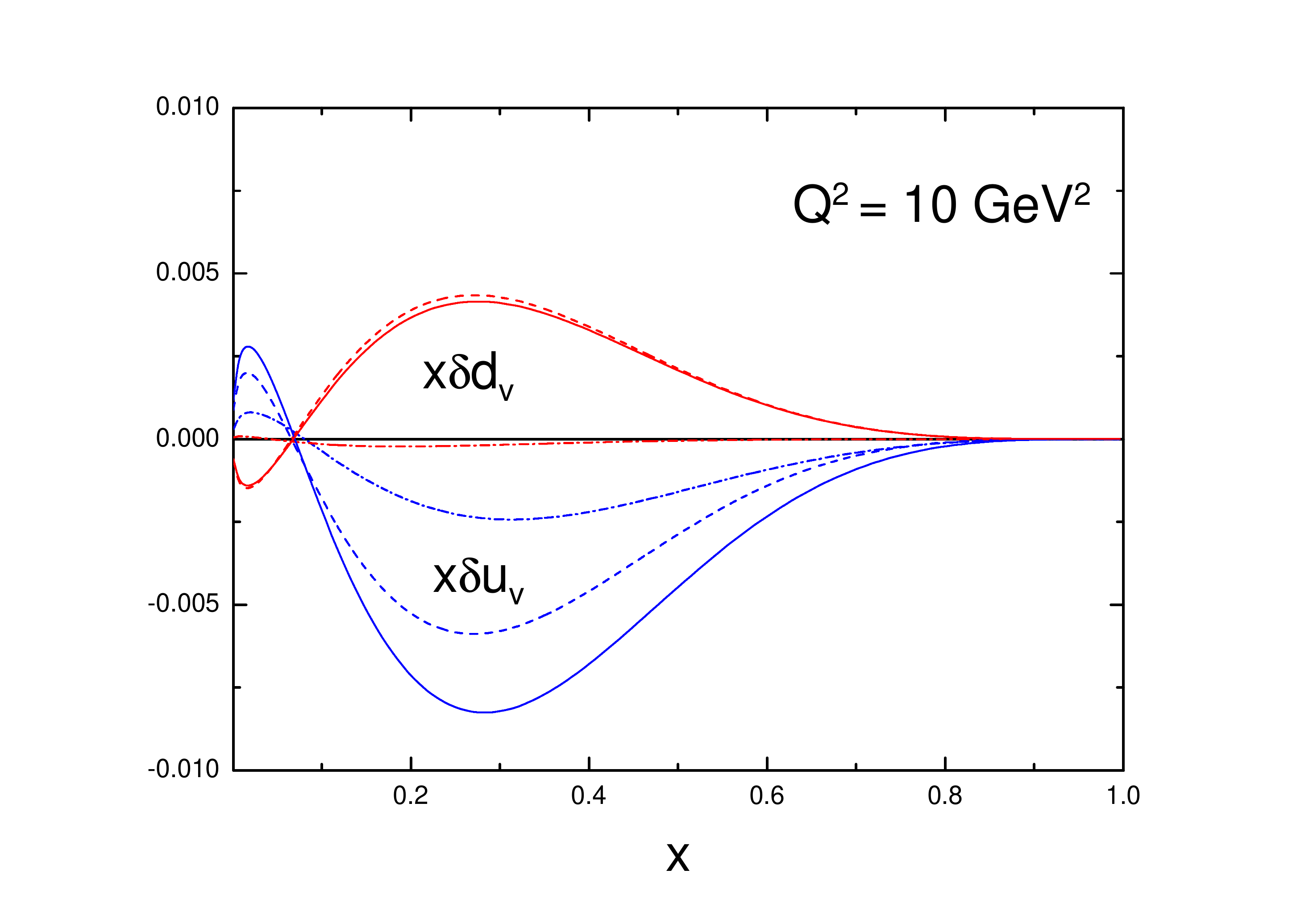}
\caption{(colour online). The isospin-violating majority $x\delta u_v$ 
and minority $x\delta d_v$ valence parton distributions 
at $Q^2=4\ \mathrm{GeV}^2$ and $10\ \mathrm{GeV}^2$. 
Dash-dotted, dashed and solid curves represent pure QED, pure QCD 
and the total contributions, respectively.}
\label{fig:csv-total}
\end{center}
\end{figure}

At $Q^2 = 4\ \mathrm{GeV}^2$, the QCD contributions to the second 
moments are derived in Ref.~\cite{Shanahan:2013} 
by extrapolating the first lattice simulations~\cite{Horsley:2010th} 
to the physical point,
\begin{equation}\label{eq:csv-lattice}
\delta U_v = - 0.0023(7)\ , \ \ \delta D_v = 0.0017(4)\ , 
\end{equation}
where the number in brackets indicates the error in the last significant figure.
These values are in good agreement with previous phenomenological 
estimates of CSV, both those calculated 
within the MIT bag model~\cite{Rodionov:1994cg,Londergan:2003pq} 
and those found in the MRST analysis~\cite{Martin:2003sk}. 
Using the simplest phenomenological parametrisation
\begin{equation}
\delta q_v(x,Q^2) = \kappa_q f(x,Q^2)\ ,
\end{equation} 
where $f(x,Q^2)$ is obtained by NLO QCD evolution from $Q_0^2 = 1\ \rm{GeV}^2$, 
and the normalisation factors are determined by 
taking the constraint, Eq.~(\ref{eq:csv-lattice}),
\begin{equation}\label{eq:parameter-kappa}
\kappa_u = - 0.26(8)\ ,\ \ \kappa_d = 0.19(4)\ . 
\end{equation}
Combining with the pure QED contributions, 
given by the lower plot of Fig.~\ref{fig:csv-coherent}, 
we show the total isospin violating distributions 
at $Q^2=4\ \mathrm{GeV}^2$ and $10\ \mathrm{GeV}^2$ 
in Fig.~\ref{fig:csv-total}.
Here we see that the influence of QED has only a small effect on the
down-quark (or minority quark) CSV. For the up (or majority) quark, we
see that the QED effects enhance the overall magnitude of the
quark-mass induced CSV.

Incorporating this new determination of partonic CSV, including the
effect of QED, we revisit the total effect of CSV in the extraction of
$\sin^2\theta_W$ by the NuTeV Collaboration.
The total correction, $\Delta s^2_W$, to $s^2_W$ arising from charge 
symmetry violation may be calculated using the very convenient 
functional, $F[s^2_W,\delta q;x]$, provided by 
the collaboration~\cite{NuTeV:2002}
\begin{equation}
\Delta s^2_W = \int_0^1 F[s^2_W,\delta q;x] x \delta q(x,Q^2) dx
\end{equation}
at the central value $Q^2 = 10\ \rm{GeV}^2$. The individual 
contributions to $\Delta s^2_W$ are listed in 
Table~\ref{tab:NuTeV-correction}. 
Therefore, the total correction arising from valence 
quark charge symmetry violation 
becomes 
\begin{equation}
\Delta s^2_W|_{\rm{total}} = \Delta s^2_W|_{\rm{QED}} 
+ \Delta s^2_W|_{\rm{QCD}} = -0.0022 \pm 0.0004 \, ,
\end{equation}
where the error is calculated by combining the errors on the 
individual contributions in quadrature. For the electromagnetic 
contribution the errors are taken as the differences between matching 
at $\mu_{\rm LO}^2$ and $\mu_{\rm NLO}^2$, while for the quark mass 
contribution the errors arise from Eq.~(\ref{eq:parameter-kappa}).
This value is consistent with that reported by Bentz et
al.~\cite{Bentz:2009yy}, namely $\Delta s^2_W = -0.0026 \pm 0.0011$,
but now with a significantly 
improved estimate of the
uncertainty associated with the QED contribution.

\begin{table}[ht]
\begin{center}
\caption{\label{tab:NuTeV-correction} The QED and QCD corrections to $\Delta s^2_W$ arising from valence quark charge symmetry violation.}
\renewcommand{\arraystretch}{1.3}
\begin{tabular}{c|rr|r}\hline
$\Delta s^2_W$   &    \multicolumn{1}{c}{$\delta u_v$}      &   \multicolumn{1}{c|}{$\delta d_v$}   &     \multicolumn{1}{c}{$\rm{Total}$} \\ \hline
        QED            &                       -0.00043(6)                        &                        0.00004(2)                     &                      -0.00039(6)                     \  \\ 
        QCD            &                       -0.00102(31)                        &                       -0.00074(17)                     &                      -0.00176(35)                     \  \\ \hline    
\end{tabular}
\end{center}
\end{table}

\section{Conclusion}
In summary, we have revisited the electromagnetic contribution
to charge symmetry violation (CSV) in the parton distribution functions
of the nucleon, which contributes the largest uncertainty associated with
the CSV correction to the NuTeV anomaly.
At very low $Q^2$ we treat the radiation of photons from the nucleon
coherently, following the suggestion of Martin and
Ryskin~\cite{Martin:2005sk}, while above the scale typically associated with
valence dominated quark models the photon emission is associated
with the individual quarks, through QED evolution~\cite{APFEL:2014}. 
The resulting electromagnetic contribution
to the combination of second moments relevant to the NuTeV anomaly,
namely $\delta D_v - \delta U_v$ is of order 0.0010 (at 10 GeV$^2$).
When used with the NuTeV functional this yields a correction of less
than 10\% of the NuTeV anomaly. Adding the latest lattice QCD estimate
of this moment~\cite{Shanahan:2013},
which is consistent with the older model dependent
calculations~\cite{Sather:1991je,Rodionov:1994cg,Londergan:1994gr},
results in a total CSV correction to $\Delta s^2_W$ of
$-0.0022 \pm 0.0004$, which constitutes a reduction in the 
anomaly of more than 40\%.
If one were to add the isovector EMC from Ref.~\cite{Cloet:2009qs}, the total
correction would be $-0.0041 \pm 0.0007$ and comparing with the quoted
anomaly, $-0.0050 \pm 0.0016$, the discrepancy with the Standard Model
appears to be resolved.
The major remaining issue is the potential asymmetry
between the $s$ and $\bar{s}$ 
distributions~\cite{Signal:1987gz,Melnitchouk:1999mv,Traini:2011tc,Mason:2007zz} 
and resolving that issue
is now of even greater importance.

\section*{Acknowledgements}
We wish to acknowledge important discussions with Ian Clo\"et and Lei Chang.
This work has been supported by the University of Adelaide and by 
the Australian Research Council through the ARC Center of Excellence for 
Particle Physics at the Terascale and through 
grants FL0992247 and DP151103101 (AWT) and DP110101265 and FT120100821 (RDY).

%
%


\end{document}